\documentclass[runningheads]{llncs}
\usepackage[T1]{fontenc}

\usepackage{graphicx}
\usepackage{multirow}
\usepackage{hhline}
\usepackage{caption}
\usepackage{subcaption}
\begin{document}
\title{Obfuscating the Hierarchy of a Digital IP}
\titlerunning{Obfuscating the Hierarchy of a Digital IP}
%
\author{Giorgi Basiashvili\orcidID{0000-0002-6021-6706} \and
Zail Ul Abideen\orcidID{0000-0002-8865-9402} \and
Samuel Pagliarini\orcidID{0000-0002-5294-0606}}
\authorrunning{G. Basiashvili et al.}
%
\institute{Centre for Hardware Security, Tallinn University of Technology (TalTech), Estonia \\
\email{\{gibasi,zain.abideen,samuel.pagliarini\}@taltech.ee}}
\maketitle              
\begin{abstract}
Numerous security threats are emerging from untrusted players in the integrated circuit (IC) ecosystem. Among them, reverse engineering practices with the intent to counterfeit, overproduce, or modify an IC are worrying. In recent years, various techniques have been proposed to mitigate the aforementioned threats but no technique seems to be adequate to hide the hierarchy of a design. Such ability to obfuscate the hierarchy is particularly important for designs that contain repeated modules. In this paper, we propose a novel way to obfuscate such designs by leveraging conventional logic synthesis. We exploit multiple optimizations that are available in the synthesis tool to create design diversity. Our security analysis, performed by using the DANA reverse engineering tool, confirms the significant impact of these optimizations on obfuscation. Among the many considered obfuscated design instances, users can find options that incur very small overheads while still confusing the work of a reverse engineer.

\end{abstract}
%
%
%
\section{Introduction} \label{sec:intro}

Security has emerged as a prime design criterion for modern integrated circuits (ICs). Alongside the continuous miniaturization trend that ICs have benefited from for decades, new security threats have emerged and are now frequently studied. Among them, the theft of intellectual property (IP) has been the target of many studies \cite{surveyLL,usenix,reverse,split_2,intro_epic,SAT,logic_5}.

The range of studied threats that ICs are susceptible to is quite large. Even before ICs undergo fabrication, the chips (or byproducts of the IC design process, e.g., netlists) can be maliciously modified, a threat that is usually termed a hardware backdoor or a hardware trojan \cite{doping}. During fabrication, an adversary located at an \emph{untrusted} foundry may proceed to analyze the IC or its components in order to gain reverse engineering knowledge \cite{reverse}. Thus, this adversary would be able to reproduce the IP for his/her own malicious purposes. This threat can take the form of a simple IP theft or IC overproduction, i.e., when the entire IC is produced beyond the contracted amount. To a large extent, the security threats herein described are due to the globalized nature of the IC supply chain; in some cases, the end-user is also considered untrusted and is assumed to be adversarial.

Many attempts have been made to counter the reverse engineering capabilities of adversaries. Approaches anchored on partial trusted fabrication have been proposed \cite{split_2} but have not been adopted by the industry. Logic locking \cite{intro_epic} studies have received a fair amount of attention and many derivatives of the original concept have emerged \cite{logic_5}. However, powerful attacks against logic locking are continuously proposed. SATisfiability-based attacks \cite{SAT}, in particular, have been very effective at breaking security assumptions of logic locking schemes.

In this paper, however, we perform a study on how to obfuscate the hierarchy of a system as not to give hints to an adversary about the intent of the system. The ability of `hiding' the hierarchy is particularly important for systems that instantiate the same module repeated times.

\section{Proposed Approach} 
\label{sec:methodology}

\subsection{Motivation and Background}

\begin{figure}[ht]
\centering
\includegraphics[width=0.92\textwidth]{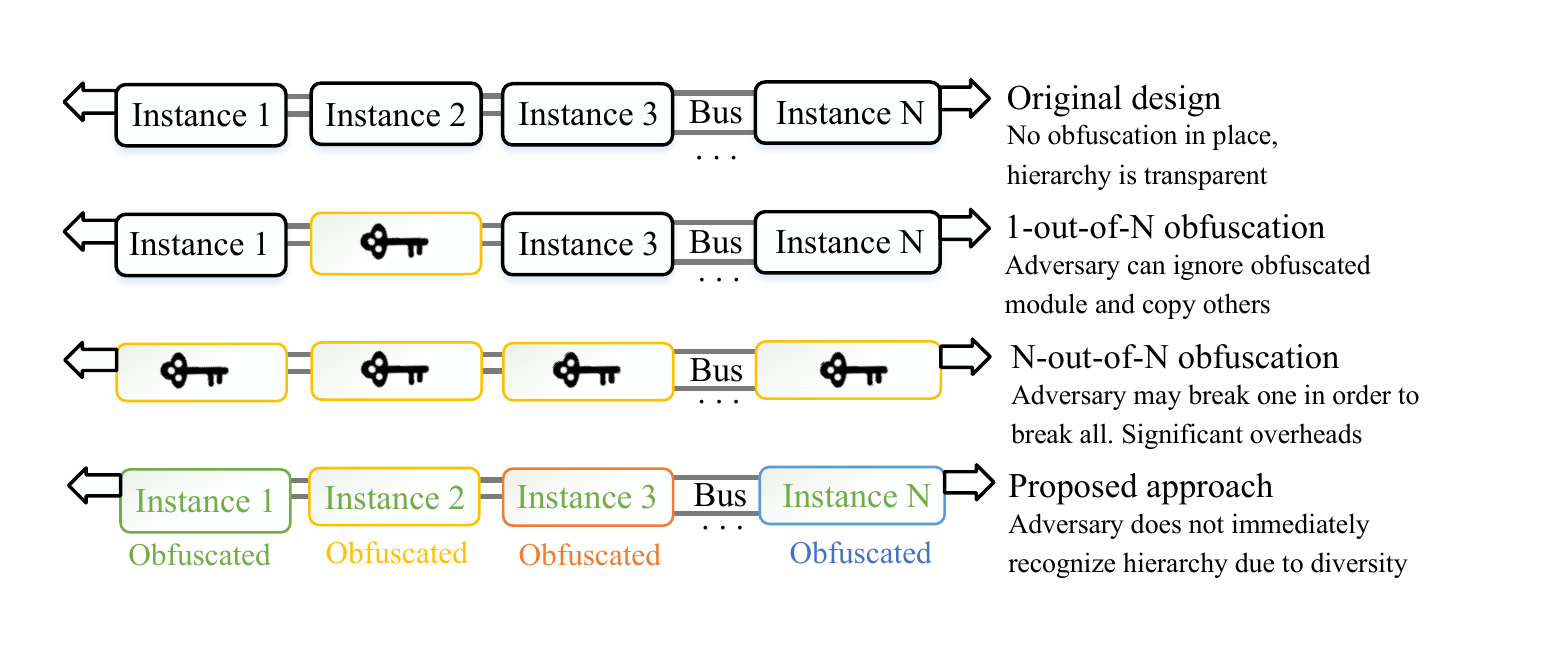} 
\caption{Approaches to obfuscate a hierarchical design, from locking to design diversity.} \vspace{-10pt}
\label{fig_diagram}
\end{figure}

There are multiple examples of designs that repeatedly instantiate the same module. For instance, in the hardware implementation of neural networks, neurons that contain multiply-and-accumulate type of functions are instantiated hundreds to thousands of times \cite{hls4ml}. This common design style is also seen in cryptographic hardware accelerators that are round-based, such as the AES \cite{aes_128}. A generic representation of such type of system is shown in Fig.~\ref{fig_diagram} (top panel), where a notion of a shared bus that connects all the repeated elements is also introduced.

\vspace{-10pt}

Next, assuming the system is an IP that is worth protecting against reverse engineering threats, one could take a state-of-the-art locking approach \cite{usenix} and apply it to a single module (second panel). While this approach seems interesting at first -- it would withstand known attacks such as SAT -- a capable adversary would bypass the problem entirely by replacing the obfuscated module with one of the transparent ones. It follows then that all instances have to be obfuscated under a key-based approach (third panel). However, even if the approach appears to have merit, once a single module is broken, they may all be broken. It is also important to notice that logic locking approaches are not overhead-free, the cost to obfuscate all $N$ modules can be rather large \cite{surveyLL}.

Generally speaking, the complete reverse engineering process is quite hard and time consuming. The process is specially hard if the adversary only has access to a finalized chip and proceeds to delaminate it, take `pictures' in steps of units of micrometers, and finally stitch them together to make sense of the design. The process is also known to be imperfect, suffering from misalignment and resolution issues. This process requires a skilled person, automation, equipment, and time. However, as said earlier, for designs that consist of N number of repeated modules, the vulnerability is higher. A small part/module of the design could leak information which prompts as a full exposure of the design to the adversary. Even if one of the submodules is poorly processed and has alignment issues, the adversary would still successfully recover a correct netlist by simply matching subnetlists of the same chip.

The illustrative example depicted in Fig.~\ref{fig_diagram} is an attempt to demonstrate that current obfuscation practices have not sufficiently tried to hide the design hierarchy. In the next subsection, we introduce a synthesis-based approach to achieve slightly modified designs in a way that would make it harder from an adversary to notice the repeated instances. The different colors on the bottom panel of the image try to convey this concept of design \textbf{diversity}.

\subsection{Hierarchical Obfuscation}

The implementation part of the ASIC design flow can be divided into two phases: logic synthesis and physical synthesis. In this work, we propose to obfuscate the hierarchy of a design during logic synthesis. In other words, we will introduce obfuscation during the process of translating an RTL description into a mapped netlist of standard cells. 

To this end, we make use of Cadence Genus. It is well know that the constraints applied during synthesis, as well as the composition of the standard cell library, play a vital role in the synthesis process. But there are many other `knobs' of the synthesis process that selectively enable certain optimization strategies. We exploit these optimization strategies to create design diversity, thus eliminating obvious regularity in a design. Furthermore, the RTL of the original design remains unchanged; users are not required to redesign their logic to make it look different. 

There are many parameters and options to define/select when performing logic synthesis, so it ought to be studied if the chosen options indeed generate different netlists. Based on the findings of \cite{tns}, we apply the following optimization techniques:

\begin{enumerate}
    \item Clock gating
    \item Ungrouping
    \item Datapath analytical
    \item Bubble pushing
    \item Tighten max transition
    \item Retiming for delay
    \item Retiming for area
    \item Clock gating + retiming for delay
    \item Bubble pushing + retiming for area
\end{enumerate}

The optimization techniques are not enabled by default, they are enabled with different attributes\footnote{For Cadence Genus, these attributes are \textit{lp\_insert\_clock\_gating}, \textit{auto\_ungroup}, \textit{dp\_analytical\_opt}, \textit{br\_seq\_in\_out\_phase\_opto}, \textit{max\_transition} and \textit{retime –min\_delay}, \textit{retime –min\_area}}. Most optimizations are enabled with a true/false setting. Exceptions are optimization 3 (it has multiple levels, from basic to extreme) and optimization 5 (it takes a time interval in picoseconds).

Even if these are classic circuit optimization techniques, we offer a short explanation for each as follows. \textbf{Clock gating} reduces the dynamic power of the design. It determines non-enabled behavior of the registers and prevents clock from propagating to them. Therefore, these registers are clock gated using an enable signal. This technique incurs a small increase in area due to the gating logic. Synthesis tools are responsible to infer enable signals automatically. \textbf{Ungrouping} allows the synthesis tool to flatten the design hierarchy and consider optimizations that traverse boundaries. This optimization typically saves area and improves timing. \textit{Datapath analytical} performs aggressive datapath optimization that compromises area for timing. When \textbf{bubble sort} is enabled, it pushes the inverters between in/out pins of flip-flops. This technique is also
referred to as an inversion of sequential (output) cells. \textbf{Tighten max transition} is related with the transition time, it is a longest time required to change the logic state. Tightening max transition results in buffers being placed on the signals with slow transitions, even for paths where this phenomenon would cause no timing violation. Max transition (clock or data) is the maximum slew, it comes either from the library or the designer manually can target in the constrained file. This helps to achieve the timing closure. \textbf{Retiming} repositions combinational cells with respect to flip-flops, from one stage to another stage. It is typically targeted for delay. It can also be issued to target area but never at the cost of delay, that is, delay remains the primary optimization target. 

In most experiments reported later on, one and only one optimization is enabled at a time. For optimizations 8 and 9, two optimizations are exploited concurrently, thus we utilize a plus sign and label them “opt\_A + opt\_B.”. For all experiments, the synthesis effort was kept medium for fairness.

\begin{figure}[ht]
\centering \footnotesize
\includegraphics[width=0.58\linewidth]{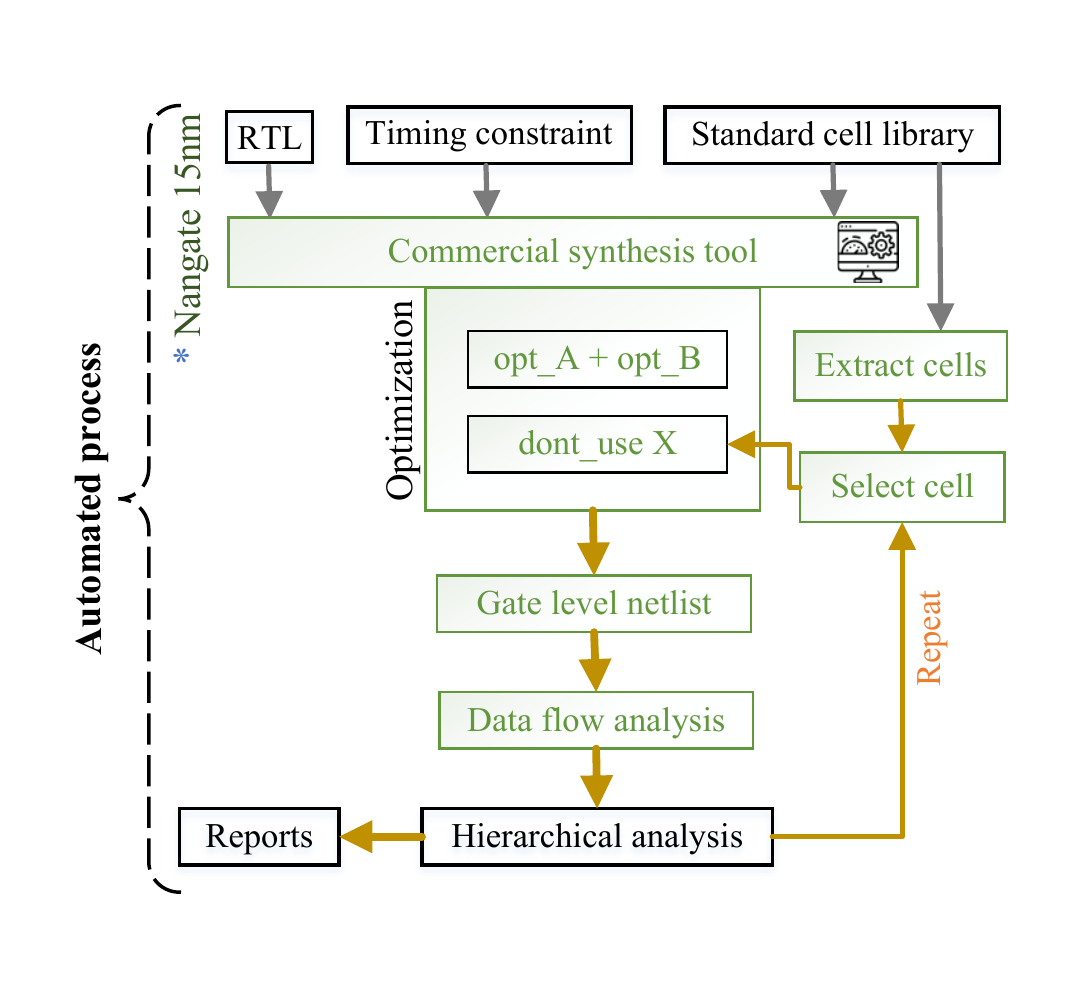}
\caption{The methodology to evaluate the hierarchy of design in the context of reverse engineering.} \vspace{-10pt}
\centering
\label{fig:meth_imp}
\end{figure}

Fig. \ref{fig:meth_imp} illustrates our methodology that exploits the aforementioned techniques during the logic synthesis to evaluate the obfuscation of the design's hierarchy. The complete process is fully automated and scripted to enable a push-button analysis. We provide RTL description (i.e., Verilog or VHDL), timing constraint and standard cell library of the targeted technology. We use the Nangate 15nm library of standard cells throughout the evaluation. 

Once the input files are provided to the commercial synthesis tool, then it generates the gate-level netlists. More precisely, the logic synthesis internally does a bit more, it synthesizes, optimizes and maps the netlist. The gate-level netlist is the mapped one exported from the synthesis tool. This process is straightforward and the Tcl script helps to achieve the automation. During the synthesis process, a part of optimization is targeted with a custom parameter as highlighted in the center of Fig \ref{fig:meth_imp}. The working principle of utilized optimization techniques and their corresponding parameters are reported in the previous section. We enable one optimization at a time (opt\_A) and then we combine another technique (opt\_A + opt\_B) in the synthesis tool.

 \vspace{-10pt}

Besides this, we exploit another technique that also transforms the structure of the design. We force the synthesis tool, with the help of \textit{set\_dont\_use} command, to avoid a given cell. Hence, the composition of the standard cell library is changed. Concerning the \textit{set\_dont\_use} command, we highlight that the Nangate library contains 67 cells. One by one, we force the synthesis not to use a single cell, synthesize the design, and export a netlist. For instance, in the first run, we prevent the $AND2\_X1$ (2-input AND with a driving strength of 1) cell from being used. In simple words, the synthesis tool will explore other available cells from the library to make a functionally equivalent $AND2\_X1$. The use of standard cells with different drive strengths or completely different cells will lead to changes in area, power, and possibly timing of the design. 
We repeat this process for all the cells, available in the library. By using this technique, every design will be functionally equivalent but marginally different (or distinct) from the previous one. First, this process was done without any optimizations, then it was repeated for the previously mentioned optimizations and their combinations. Next, we use the \textit{set\_dont\_use} technique to explore the synthesis space of the design. The complete process generates 603 netlists which we calculate with the number of cells multiplied by optimization techniques ($67 \times 9$). 

However, in some cases, this difference might not be enough to alter the design significantly enough. Accordingly, we have to check every design and verify its characteristics, such as area, power and delay of the critical path to eliminate the duplicate designs. This is achieved with a custom script that shrinks the netlist and keeps only unique instances.


In the next step, we exploit an open-source tool named DANA to analyze the design \cite{DANA}. DANA stands for Dataflow Analysis for Gate-Level Netlist Reverse Engineering. DANA is a fully automated, technology-agnostic data flow methodology for gate-level netlists. It analyses the individual FFs and groups them into high-level registers. It provides an easily readable summary, such that the user of tool can make sense of a netlist that \emph{a priori} looks like a sea of gates. We perform data flow analysis for each individual netlist and export a report. We repeated this process for all the unique netlists to conclude the analysis. For collecting final results and statistics of this analysis, we use another custom script which we present in the next section.

\section{Results} \label{sec:results}
This section reports the results of our proposed methodology. Recalling again, the objective was to develop a key-less and structural obfuscation methodology that will apply to circuits that have modules instantiated multiple times. 

To evaluate our methodology, we have selected a representative design that displays such pattern. We have selected a Global Positioning System (GPS) correlator, which is one of the integral parts of a GPS-capable hardware. A correlator attempts to identify to which satellite of the GPS constellation it is talking too. A system with multiple correlators executes this tasks in parallel, yielding in faster signal acquisition and therefore short sync time. 


We used the RTL description of the GPS correlator and generated the results for the a single design. We did not change the RTL of the design throughout the analysis for fairness. We note the number of unique netlists that are generated by optimization techniques. A total of 509 unique designs were generated. With \textit{set\_dont\_use cell}, it was able to generate 55 designs. Clock gating, datapath analytical and bubble pushing generates 50, 56 and 53 designs each. Most of the unique designs were generated by tightening max transition (62), retiming for the delay (61), and its combination with clock gating (62). However, it should be noted that ungrouping was not able to generate a single unique design. Retiming for area and its combination with bubble pushing generates 53 and 57 designs. Nevertheless, the number of duplicate designs varies depending on the optimization strategy.


\subsection{Power-Performance-Area evaluations} \label{sec:power_performance_area}
Our proposed obfuscation is key-less and infers almost a little overhead or almost zero. The performance does not impact the optimization techniques. But, the area and power vary therefore it should be investigated. We have used a very relaxed clock frequency in order to allow the synthesis tool to make less constrained decisions. Table \ref{tab:min_max} shows the minimum and maximum values for the area and power trade-offs with respect to changes in the applied optimization technique. The first column list the applied optimization technique, columns two to three for the minimum and maximum values of the area, columns four to five for the minimum and maximum values of the leakage power and the last two columns list the values for the dynamic power. 

\begin{table}[ht]
\centering
\resizebox{\textwidth}{!}{
\begin{tabular}{|l|l|l|l|l|p{1.8cm}|l|p{1.8cm}|p{0.2cm}|}
\hline
\multirow{2}{*}{\textbf{Optimization technique}} & \multicolumn{2}{|c|}{\textbf{Area ($\mu m^2$)}} & \multicolumn{2}{c|}{\textbf{Cells}} & \multicolumn{2}{c|}{\textbf{Leakage power ($mW$)}} & \multicolumn{2}{c|}{\textbf{Dynamic power ($mW$)}} \\ \hhline{~--------}
& \textbf{Min} & \textbf{Max} & \textbf{Min} & \textbf{Max} & \textbf{Min} & \textbf{Max} & \textbf{Min} & \textbf{Max} \\ \hline
\textbf{Baseline} & 432.9 & 461.6 & 762 & 845 & 0.012 & 0.013 & 2.368 & 2.414 \\ \hline
\textbf{Clock Gating} & 432.7 & 462.2 & 750 & 810 & 0.012 & 0.013 & 0.600 & 0.935 \\ \hline
\textbf{Ungrouping} & 432.9 & 461.6 & 762 & 845 & 0.012 & 0.013 & 2.368 & 2.414 \\ \hline
\textbf{Datapath Analytical} & 433.2 & 458.8 & 750 & 821 & 0.012 & 0.013 & 2.368 & 2.431 \\ \hline
\textbf{Bubble Pushing} & 437.1 & 459.1 & 758 & 840 & 0.012 & 0.013 & 2.293 & 2.469 \\ \hline
\textbf{Tightening max transition} & 598.8 & 990.3 & 1168 & 1644 & 0.022 & 0.062 & 0.706 & 2.091 \\ \hline
\textbf{Retiming for Delay} & 440.9 & 600.9 & 737 & 1047 & 0.012 & 0.018 & 2.944 & 4.398 \\ \hline
\textbf{Retiming for Area} & 434.7 & 458.8 & 767 & 840 & 0.012 & 0.013 & 2.399 & 2.460 \\ \hline
\textbf{Clock Gating + Retiming for Delay} & 425.2 & 560.9 & 711 & 966 & 0.012 & 0.018 & 1.084 & 1.562 \\ \hline
\textbf{Bubble Pushing + Retiming for Area} & 440.0 & 460.3 & 755 & 842 & 0.012 & 0.013 & 2.350 & 2.537 \\ \hline
\end{tabular}
}
\caption{Minimum and Maximum values of area, number of cells, leakage and dynamic power of the generated designs} \vspace{-10pt}
\label{tab:min_max}
\end{table}

\vspace{-10pt}

\begin{figure}[ht]
\centering
\includegraphics[width=0.85\textwidth]{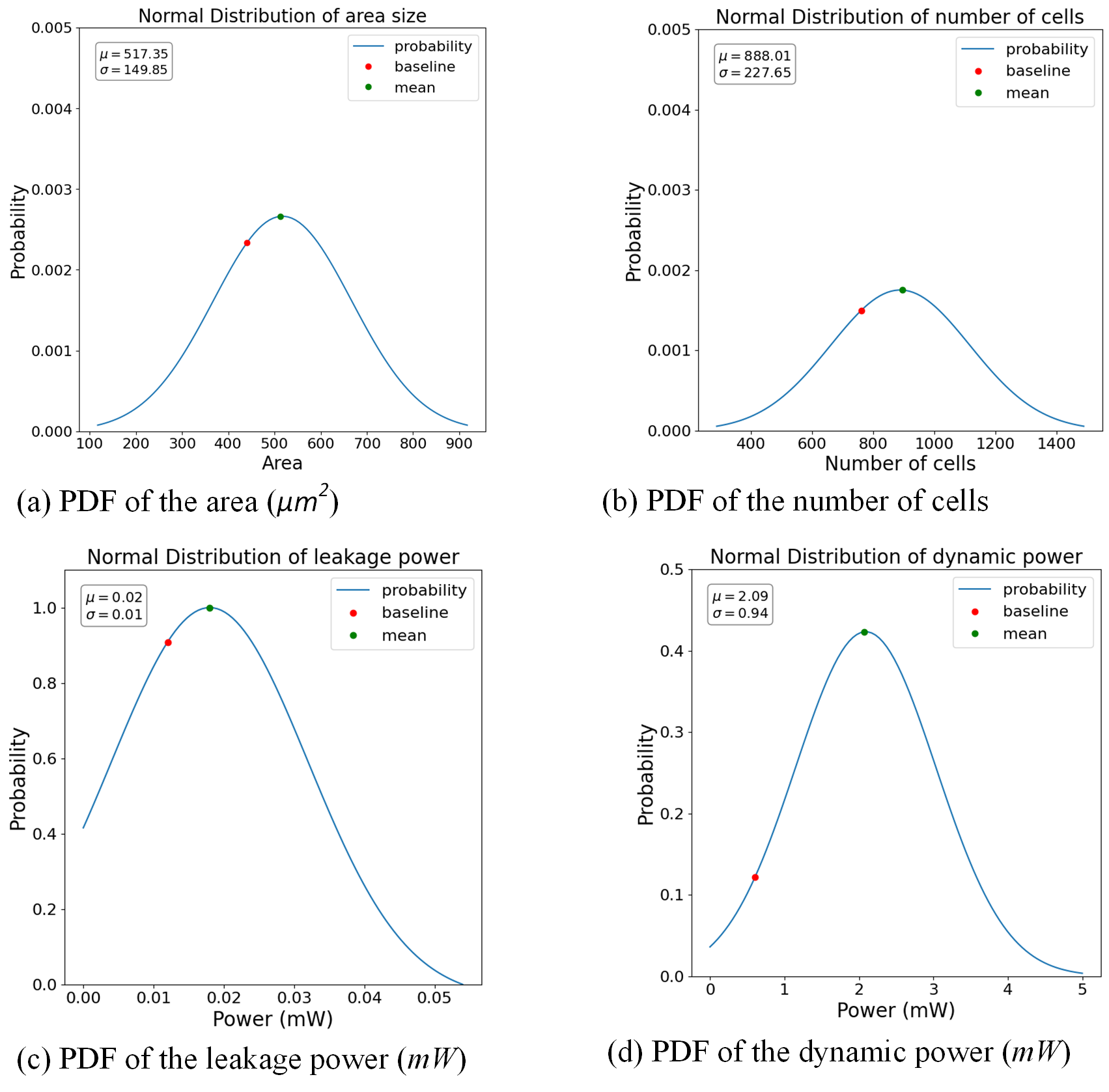} \vspace{-10pt}
\caption{PDFs in terms of area, number of cells, leakage and dynamic power for unique designs.} \vspace{-10pt}
\label{fig:PDFs_of_designs}
\end{figure}

Fig. \ref{fig:PDFs_of_designs} illustrates the probability distribution function (PDF) for the unique netlist. Panels (a), (b), (c) and (d) highlight the PDFs for the area, number of cells, leakage power and dynamic power, respectively. 
The red dot represents the location of the baseline design, and gives us an idea about the overall results. The data of the distinct netlists shows the perfect fit for the normal distribution curves. The distribution of the leakage power is a little bit different but still faces variance. Thus, we have a wide range of variety which indicates that there are significantly different designs from one another. Concerning panel (a), the baseline design is approximately closer to the mean value. Almost half of the designs have less area as compared to the baseline design. The same case is for the number of cells as seen in panel (b). Similarly, the leakage power of the baseline design is closer to the mean value. More than half of the designs consume more leakage power as compared to the baseline design. Regarding the dynamic power, the baseline design is far from the mean value and a large number of designs consume higher power as compared to the baseline design. It is noteworthy, that we observe the change in the hierarchy of the structure and the effect of the variation is reflected in the area, number of cells, leakage power and dynamic power.

Next, we are going to observe the percentage increase and decrease of the area, number of cells, leakage power and dynamic power. Table \ref{tab:statistics} lists the analysis of different overheads for their corresponding techniques. The first column list the optimization technique, the second column shows the percentage increase/decrease in the area, the third column shows the percentage increase/decrease in the number of cells, last two column represents the leakage and dynamic power. 

\begin{table}[ht]
\centering 
\resizebox{\textwidth}{!}{
\begin{tabular}{|l|l|l|l|l|l|l|l|}
\hline
\textbf{Optimization technique} & \textbf{Area (\%)} & \textbf{Cells (\%)} & \textbf{Dynamic Power (\%)} & \textbf{Leakage Power} \\ \hline
\textbf{Clock gating}                       & -0.3 & -0.7 & -119.4    & 0           \\ \hline
\textbf{Ungrouping}                         & 0      & 0      & 0           & 0           \\ \hline
\textbf{Datapath analytical}                & -1.4  & -2.9  & -0.4      & 0           \\ \hline
\textbf{Bubble pushing}                     & -0.5  & -0.7 & -2.7      & 0           \\ \hline
\textbf{Tighten max transition}            & +69.8  & +60.1  & -84.3     & +131.4       \\ \hline
\textbf{Retiming for delay}                 & +20.1  & +12.7  & +51.3      & +28.5       \\ \hline
\textbf{Retiming for area}                  & -0.5 & -0.3 & +0.7        & 0           \\ \hline
\textbf{Clock gating + Retiming for delay}  & +18.9   & +13.6  & -66.5      & +34.4       \\ \hline
\textbf{Bubble pushing + Retiming for area} & +0.4   & -2.1  & +4.5        & 0           \\ \hline \end{tabular}
} 
\caption{Percent increase/decrease in the baseline design and a variants generated with the corresponding optimization technique} \vspace{-10pt}
\label{tab:statistics}
\end{table}
\vspace{-10pt}

We note that clock gating offers a significant decrease in the dynamic power. Datapath analytical lowers area, cells and dynamic power between 1-3\%. The same is happening for the bubble pushing. Tighten max transition has a significant impact on area, cells and leakage power. But it shows a remarkable decrease in the dynamic power (84.3\%). We should note that a large number of distinct designs were generated from this technique. The retiming for delay also has a similar behavior for area and cells (20.18\% and 12.71\% increase) but it also shows an increase in the leakage and dynamic power. The combination of clock gating and retiming for delay shows an increase in every parameter except dynamic power, analogous to tighten max transition. The combination of bubble pushing and retiming for the area also shows a little increase/decrease in the parameters. In a nutshell, all the techniques have little impact on the area, cells and power consumption except tighten max transition, retiming for delay and a combination of clock gating \& retiming for delay. 

\subsection{Security analysis} \label{sec:security}

This section details the security analysis with DANA. The tool analyses the dataflow between flip-flops to structure the registers. From the abstraction level, the entire design contains flip-flops, connections to their respective sequential successors and predecessors. The constructed relationship between flip-flops helps the adversary to accomplish his/hers reverse engineering goals. DANA offers two modes: (a) Normal Mode and (b) Steered Mode. In Normal Mode, DANA autonomously analyzes the given netlist, without any prioritization. Using the Steered Mode, the analyst can alternatively take care of extra information to virtually ``steer'' the algorithms. This includes prioritizing DANA for the specific register sizes, i.e., a reverse engineer learns different information from datasheets. Furthermore, If the reverse engineer already has a clue that the design under analysis is, for instance, a 16-bit CPU, he would hope to observe multiple 16-bit register and thus steer DANA towards said size. 

\subsubsection{Evaluation method} \label{subsub:eval_method}
\begin{figure}[ht]
\centering
\includegraphics[width=0.85\textwidth]{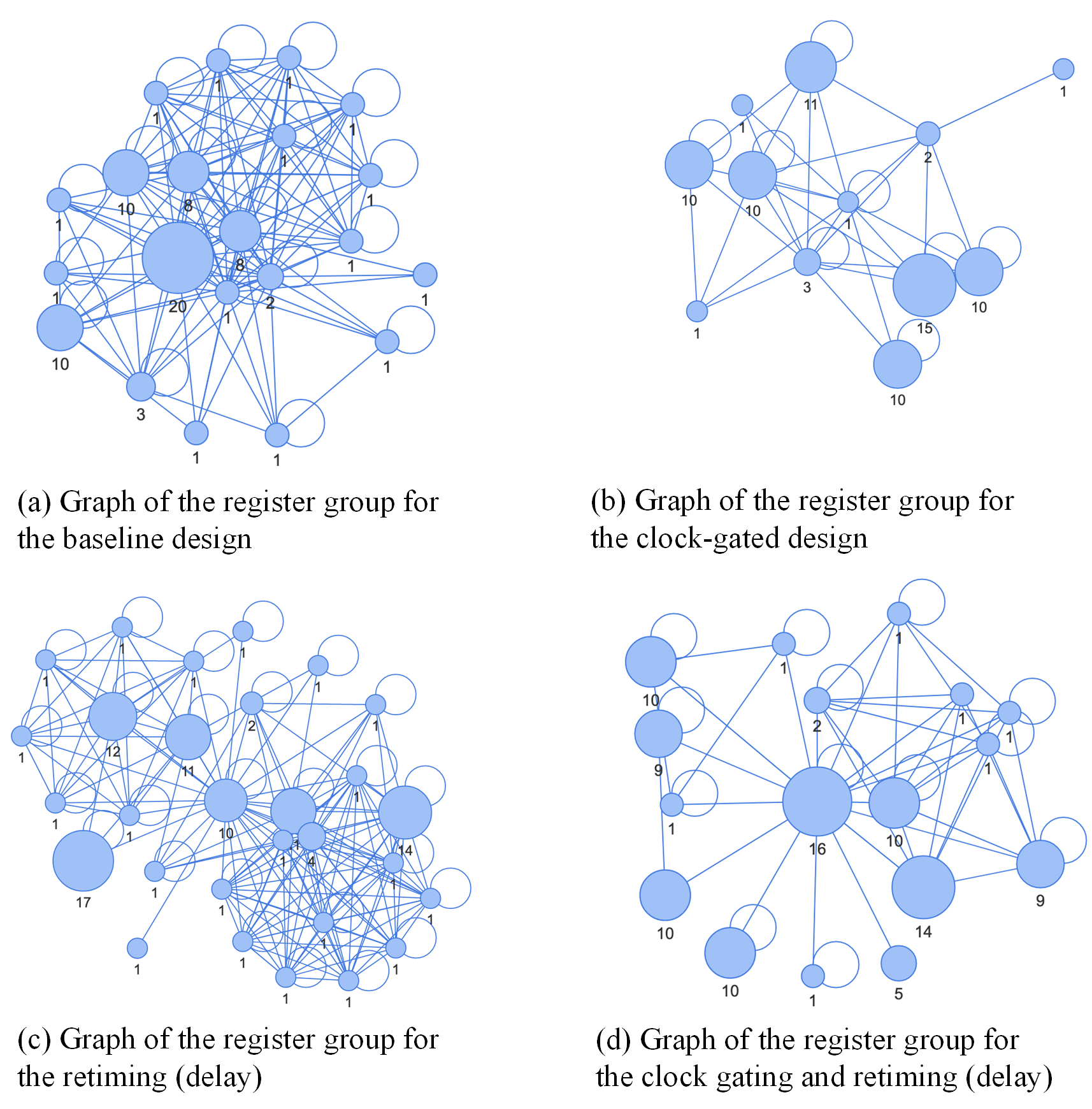} \vspace{-10pt}
\caption{Graphs of the register group for various optimization techniques).} \vspace{-10pt}
\label{fig:graphs_DANA}
\end{figure}

To accomplish the patterns for the reverse engineering, we first executed DANA in a normal model and then in the steered mode. We run DANA on a a machine equipped with 32 processors (Intel(R) Xeon(R) Platinum 8356H CPU @ 3.90GHz). After running DANA twice, we compare the results with the baseline design to analyze how the hierarchy is varying with respect to the applied technique. 
Fig \ref{fig:graphs_DANA} shows the register groups for the baseline and various optimization techniques. The figure in panel (a) shows the baseline design that includes different sizes of registers. Blue bubbles show the registers, straight lines are connections between different registers, and circular lines are connections from a register to itself. Concerning the size of registers, the larger the bubble is, the larger the size of the register. From the graph, it is clear that it shows two 10-bit registers and other sizes of registers. This hierarchy shows the internal register size and connectivity of the registers. This is straight forward clue for the adversary to restructure the registers and its connected circuitry. 

Now, we investigate how optimization changes the characteristics of the design. The figure in panel (b) shows the synthesized design where we enabled clock gating optimization in commercial synthesis tool. We can see different 10-bit registers. A large number of register are diminished and we only see few registers with a variety of sizes. It allows us to clearly see distinctions between different groupings. In these unrolled designs, the sizes of register does not correspond to the correct sizes declared in the RTL code. Panel (c) of Fig. \ref{fig:graphs_DANA} shows the graph for the synthesized design with retiming for delay. Here, we can analyze that complete graph consists of a large number of register and we still can see a 10-bit register. This also offers a unique structure of the design and DANA is unable to map it in the same way. In the next example, shown in panel (d), we exploit two different optimizations (clock-gating and retiming for delay) at the same time. Again, we obtain a distinct graph. To summarize these results, we can confidently state that a design composed of many instances of the same module but each instance is synthesized differently, will present itself as a \textbf{challenge to a reverse engineering adversary}. 

\begin{figure}[ht]
\centering
\includegraphics[width=0.40\textwidth]{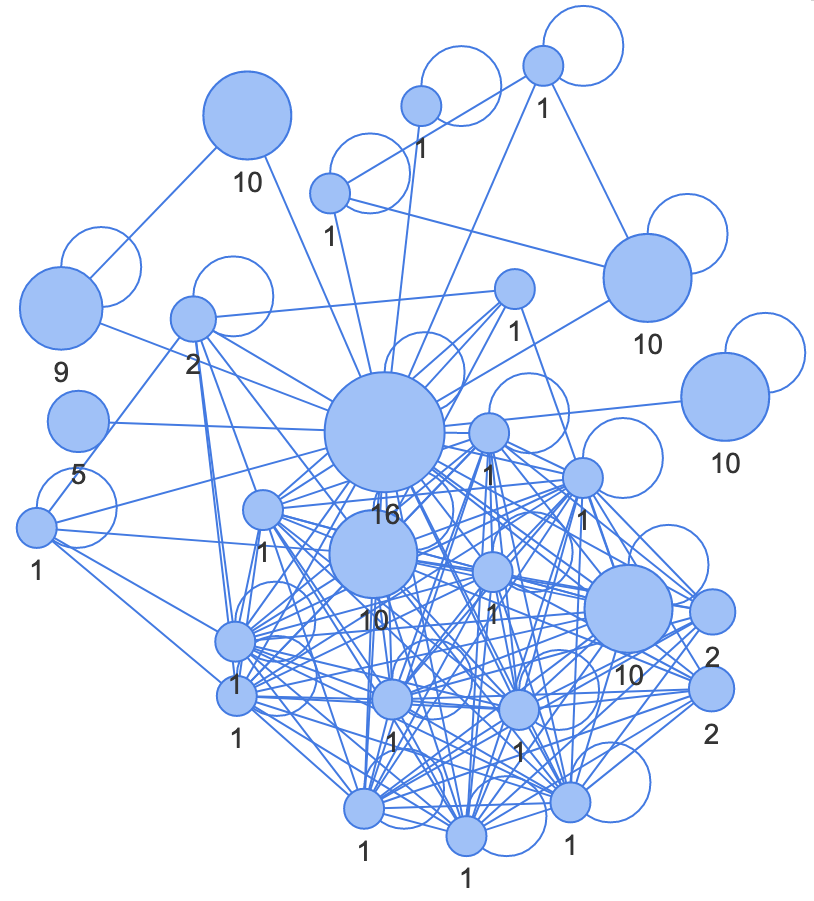} \vspace{-10pt}
\caption{Graph of the register group for the clock-gating and retiming (delay) with steered mode (Register size 10).} \vspace{-10pt}
\label{fig:steered_dana_10}
\end{figure}

All these experiments presented so far were executed in the normal mode of DANA. Now, we exploit the steer mode of DANA with the register size of 10-bits as shown in Fig. \ref{fig:steered_dana_10}. It is a fair assumption that from non-steered mode, and adversary might reach the conclusion that 10-bit registers are present. We can see that the structure of the design is explicitly different from the previous ones. DANA still is unable to highlights clues even in steering mode. This implied the high level of obfuscation for the design. The adversary make use of the different reverse engineering tools along with high skills but still it requires an additional effort to correctly identify the design. Our applied optimization techniques perfectly modifies the structure of the design. This places barriers for DANA's clustering algorithm which incorrectly identifies the register group. This is the case for every optimization technique. 

\subsubsection{Remarks} \label{subsub:remarks}
The optimizations techniques are contributing towards obfuscation, to confuse the adversary to understand the architecture of design. It is noteworthy that the combination of two different  optimizations has a large impact on the hierarchy of the design (see Fig. \ref{fig:graphs_DANA}, panels (c) and (d)). On the other hand, if the design goal is keep the overheads under control, then tighten max transition and retiming for delay are excluded for the list. However, many other optimizations still remain attractive solutions. 

\section{Conclusion}
Numerous designs consist of repeated modules or multiple copies for a specific part of the design. These designs are highly vulnerable to reverse engineering. In this paper, we have presented a unique obfuscation by leveraging the logic synthesis and different optimizations techniques. Our methodology specifically targets these types of designs. Our proposed flow is beneficial as the crucial initial step for security analysis. Our flow for the obfuscation is completely automated and does not incur high overheads, nor RTL changes. Our extensive analysis of a large number of designs reveals that optimization techniques partially or completely modify the structural representation of the design, thus creating a challenge for reverse engineering.

\subsubsection{Acknowledgements} This work has been partially conducted in the project ``ICT programme'' which was supported by the European Union through the ESF.

%
%
%
\bibliographystyle{splncs04}
\bibliography{obfuscation}

\end{document}